\documentclass[aps,prd,twocolumn,groupedaddress,preprintnumbers,nofootinbib]{revtex4}

\usepackage{url}

\usepackage[latin1]{inputenc}
\usepackage[english]{babel}

\usepackage{amssymb,amsmath,amsthm,cancel,hyperref,graphicx,xcolor}
\usepackage{picinpar,graphicx,xypic}
\usepackage{mathrsfs}
\usepackage[a4paper, top=2.2cm, bottom=2.2cm, left=2cm, right=2cm, bindingoffset=0mm]{geometry}
\usepackage{flushend}
\usepackage{xspace}

\numberwithin{equation}{section}

\newcommand{\as}{\alpha_s}

\let\originalleft\left
\let\originalright\right
\renewcommand{\left}{\mathopen{}\mathclose\bgroup\originalleft}
\renewcommand{\right}{\aftergroup\egroup\originalright}

\def\beq{\begin{equation}}
\def\eeq{\end{equation}}
\def\({\left(}
\def\){\right)}
\def\[{\left[}
\def\]{\right]}
\allowdisplaybreaks

\graphicspath{{images/}}

\newcommand\Matrix{{\sc Matrix}\xspace}
\newcommand\lpc{\mbox{linPCs}\xspace}

\begin{document}

\title{ Linear power corrections for two-body kinematics in the \boldmath{$q_T$} subtraction formalism  }

\author{Luca Buonocore}
\affiliation{Physik Institut, Universit\"at Z\"urich, CH-8057 Z\"urich, Switzerland}
\author{Stefan Kallweit}
\affiliation{Dipartimento di Fisica, Universit\`{a} degli Studi di Milano-Bicocca and INFN, Sezione di Milano-Bicocca, I-20126, Milan, Italy}
\author{Luca Rottoli}
\affiliation{Physik Institut, Universit\"at Z\"urich, CH-8057 Z\"urich, Switzerland}
\author{Marius Wiesemann}
\affiliation{Max-Planck-Institut f\"ur Physik, 80805 M\"unchen, DE-80805 Germany}

\preprint{ZU-TH 57/21, MPP-2021-198 }

\begin{abstract}

Transverse-momentum cuts on undistinguished particles in two-body final states induce an enhanced sensitivity to low momentum scales. 
This undesirable feature, which ultimately leads to an instability of the fixed-order series, poses additional challenges to non-local subtraction schemes.
In this letter, we address this issue for general colour-singlet processes within the $q_T$-subtraction formalism, focussing on neutral-current Drell--Yan production.
We present a simple procedure to reduce the dependence on the slicing parameter from linear to quadratic, by accounting for the linear power corrections through an appropriate recoil prescription.
We observe a dramatical improvement of the numerical convergence and a reduction of the systematic uncertainties.
We also discuss how a linear dependence in $q_T$ can be avoided for Drell--Yan production by using staggered cuts, which, to the best of our understanding, could be used in experimental analyses.
We show that our approach can be successfully applied also to on-shell $ZZ$ production.
We finally study diphoton production and verify that our approach is insufficient to capture the linear power corrections introduced by the isolation \mbox{procedure. The recoil prescription is available in version 2.1 of \Matrix.}

\end{abstract}

\maketitle

Accurate comparisons between experimental measurements and theoretical predictions are a key ingredient of the precision programme at the Large Hadron Collider (LHC).
In order to minimize model-dependent assumptions as a source of bias in data--theory comparisons, experimental analyses at high energy colliders define fiducial regions close to the phase space accessible to the experiments.   
Within the fiducial phase space theoretical predictions can be compared directly with data without relying on models to extrapolate beyond the experimental acceptance.

The definition of the fiducial phase space translates to a set of cuts on kinematical variables of the detected particles, which typically involve their transverse momenta and (pseudo-)rapidities.
In two-body final state systems, a lower limit on the trasverse momenta of final state particles is usually applied. Typical choices in experimental analyses are the application of a common value for the minimum transverse momentum (\textit{symmetric} cuts, henceforth) or a  different value of the transverse momenta of the leading and subleading final state particles (\textit{asymmetric} cuts). Other choices of cuts on the minimum transverse momentum are possible, although much less common; for instance, different cuts could be applied on identified particles in the final state, e.g. on the positive and negative leptons in neutral current Drell--Yan production (\textit{staggered} cuts).

It was pointed out more than two decades ago~\cite{Klasen:1995xe,Harris:1997hz,Frixione:1997ks} that the use of a common minimum transverse momentum cut on each particle in a two-body final state can spoil the convergence of the fixed-order series.
This instability of the perturbative series is due to an enhanced sensitivity to soft radiation when the two particles are back-to-back in the transverse plane, which manifests itself in the form of large logarithmic contributions 
of the imbalance.
Although initially pointed out in the context of dijet production, the poor behaviour of the perturbative series in the presence of symmetric cuts affects also other relevant collider processes, such as neutral-current Drell--Yan production or the two-body decays of a Higgs boson.

The enhanced sensitivity to soft radiation when symmetric cuts are applied poses a challenge~\cite{Grazzini:2017mhc,Catani:2018krb,Ebert:2019zkb,Alekhin:2021xcu} to non-local subtraction methods, such as $q_T$-subtraction~\cite{Catani:2007vq} or $N$-jettiness subtraction~\cite{Gaunt:2015pea,Boughezal:2015dva,Boughezal:2015eha}.
In the context of $q_T$ subtraction for colour-singlet production the problem is related to the fact that the scaling of missing power corrections is changed from being quadratic~\cite{Grazzini:2016ctr,Ebert:2018gsn,Buonocore:2019puv,Cieri:2019tfv,Oleari:2020wvt} to linear in $q_T$~\cite{Grazzini:2017mhc,Ebert:2019zkb}.
In order to correctly compute perturbative corrections, one should ideally lower the technical slicing cutoff to very small values and/or perform an extrapolation to a vanishing cutoff, which affects stability and performance of these methods especially at higher orders.
This situation challenges the applicability of these subtraction methods for benchmark processes like neutral-current Drell--Yan production, where symmetric cuts have been used in the past and a particularly high precision is demanded.

As a consequence of the observations made in Refs.~\cite{Klasen:1995xe,Harris:1997hz,Frixione:1997ks}, experimental analyses started to define fiducial regions by applying asymmetric  cuts on the transverse momenta of the leading and subleading final-state particles (ordered in transverse momentum) in processes with a two-body final state.
The use of asymmetric cuts is now common practice in the definition of the fiducial phase space, for instance in recent 
\mbox{$H \to \gamma \gamma$} analyses at the LHC~\cite{ATLAS:2018hxb,CMS:2021kom}. 
Nevertheless, symmetric cuts are still in use in various experimental analyses, most notably
in some neutral-current Drell--Yan measurements, see e.g.\ Refs.~\cite{ATLAS:2019zci,CMS:2019raw}.

However, relying on asymmetric cuts in general does not cure the problem of linear power corrections (\lpc) present in the symmetric case.
Indeed, the same linear dependence is observed when asymmetric transverse-momentum cuts are applied on the leading and subleading leptons in neutral-current Drell--Yan production or on the two final-state photons from \mbox{$H\to \gamma \gamma$} decays in the gluon fusion production channel~\cite{Grazzini:2017mhc,Ebert:2020dfc,Salam:2021tbm}.

It was recently shown that the presence of a linear dependence in $q_T$ is actually a more fundamental problem, as it ultimately leads to a factorial growth of the coefficients in the perturbative series~\cite{Salam:2021tbm}.
Despite the asymptotic limit of the series being well defined, the sequence of its fixed-order truncations is badly behaved and will eventually start to develop a divergent trend.
Therefore, there is an associated ambiguity in the fixed-order prediction, which also depends on the Casimir scaling of the processes, and can be already observed in the \mbox{$H\to \gamma \gamma$} case~\cite{Chen:2021isd,Billis:2021ecs}. 

We remark that this undesired behaviour does not depend on the subtraction method used. It can be prevented by using alternative fiducial cuts, as those suggested in Refs.~\cite{H1:1998vns,Carli:1998zr,H1:2000bqr,Rubin:2010xp,Salam:2021tbm}. 
For the Drell--Yan case, a very simple alternative is to impose different cuts on the transverse momenta of the lepton and the anti-lepton, i.e.\ the aforementioned staggered cuts.
At variance with symmetric and asymmetric cuts, these cuts do not induce a linear dependence in $q_T$, as noticed in Ref.~\cite{Grazzini:2017mhc}, and they are experimentally viable thanks to the excellent identification performance of the experimental apparatus at the LHC~\cite{ATLAS:2019qmc,CMS:2020uim}.

However, while future analyses will hopefully adopt new definitions of fiducial cuts that are free from the issues discussed above, theoretical predictions must be provided for legacy analyses that used symmetric or asymmetric cuts.
A possible option is to supplement fixed-order predictions with resummation, which stabilises the perturbative series~\cite{Salam:2021tbm,Billis:2021ecs} by including the \lpc at all orders in $\as$~\cite{Ebert:2020dfc}.
While theoretically this is probably the cleanest option, for benchmark processes like Drell--Yan production it would still be desirable to have predictions at fixed order, which are required, for instance, in the extraction of parton densities.

In this work, we present a simple algorithm to include the missing \lpc below the cutoff in the $q_T$-subtraction formalism of Ref.~\cite{Catani:2007vq}, circumventing the numerical instabilities related to the use of a tiny value of the slicing parameter.
In this way, the missing power corrections become quadratic, analogously to the inclusive case, rendering the method more efficient and suitable for benchmarking purposes.
The idea is based on the observation that the origin of \lpc is ultimately related to phase-space effects,
which was used to compute the \lpc{} in $q_T$ subtraction for Higgs and Drell--Yan production in Refs.~\cite{Billis:2021ecs,Ebert:2020dfc}.

We start by recalling the formula for the cumulative cross section computed in the $q_T$-subtraction formalism~\cite{Catani:2007vq}, which can be written as 
\begin{widetext}
\begin{equation}\label{eq:qtsub}
\sigma^{q_T\textrm{-sub}}(r_{\rm cut}) = \int d \Phi_F  \mathcal H + \left[ \int d \Phi_{F+{\rm jet}} \frac{d \sigma^{F+{\rm jet}} }{d \Phi_{F+{\rm jet}}} \theta(q_T/Q-r_{\rm cut}) - \int d \Phi_F \int d q_T  \frac{d\sigma^{ \rm CT}}{d \Phi_F dq_T} \theta(q_T/Q-r_{\rm cut})\right] ,\tag{1}
\end{equation}
\end{widetext}
where  the hard-virtual function $\mathcal H$ is independent of the transverse momentum $q_T$ of the colour-singlet system $F$ and defined on the Born phase space $\Phi_F$. 
The second term corresponds to the cross section for $F$+jet production with
the respective phase space denoted as $\Phi_{F+{\rm jet}}$, while the $q_T$-subtraction
counter term (CT) includes all contributions that are singular in the limit $q_T \rightarrow 0$ and 
is computed from the expansion of the $q_T$-resummation formula to the given 
fixed order in $\as$.
As a consequence, the difference between the second and the third term in Eq.~\eqref{eq:qtsub} contains only non-singular contributions in $q_T$. 
However, since both the $F$+jet cross section and the (non-local) subtraction term diverge at small $q_T$, a $q_T$-slicing cutoff must be imposed in order to numerically compute the quantity in square brackets. 
Typically, a cutoff $r_{\rm cut}$ is introduced on the dimensionless quantity $q_{T}/Q$, where $Q$ is the hard scale of the process.

Due to the presence of this slicing cutoff the cross section in Eq.~\eqref{eq:qtsub} misses non-singular contributions below $r_{\rm cut}$.
While some work has been devoted to study such corrections in the inclusive case~\cite{Ebert:2018gsn,Buonocore:2019puv,Cieri:2019tfv,Oleari:2020wvt}, an exact computation for general processes in presence of fiducial cuts is more challenging.
In Ref.~\cite{Ebert:2020dfc} the authors performed an all-order resummation of \lpc for Drell--Yan production, using a tensor decomposition of the hadronic and leptonic tensors, and showed that this is equivalent to resorting to a suitable recoil prescription as applied in the context of $q_T$ resummation~\cite{Catani:2015vma,Grazzini:2015wpa,Camarda:2019zyx,Becher:2019bnm,Scimemi:2019cmh,Bacchetta:2019sam,Becher:2020ugp,Re:2021con,Ju:2021lah}.
In particular, the \lpc can be resummed to all orders in perturbation theory by boosting the leading-order kinematics to a frame in which the colour-singlet system has transverse momentum $q_T$~\cite{Ebert:2020dfc}.

If such recoil prescription is implemented, also the expansion of the $q_T$-resummed result captures all the \lpc to a given order in $\as$.
As a consequence, the missing \lpc below the $q_T$-slicing cutoff $r_{\rm cut}$ can be included by computing the difference
\begin{widetext}
\begin{equation}\label{eq:mainlpc}
\Delta\sigma^{\rm \lpc} (r_{\rm cut}) = \int d \Phi_F \int_{\epsilon}^{r_{\rm cut}} d r' \left(\frac{d\sigma^{ \rm CT}}{d\Phi_F dr' } \Theta_{\rm cuts} (\Phi_F^{\rm rec}) - \frac{d\sigma^{ \rm CT}}{d \Phi_F dr'} \Theta_{\rm cuts} (\Phi_F) \right),\tag{2}
\end{equation}	
\end{widetext}
where $\Theta_{\rm cuts}(\Phi)$ collects the fiducial cuts on the phase space $\Phi$, and \mbox{$\Phi_F^{\rm rec} \equiv \Phi_F^{\rm rec} (\Phi_F,r')$} is the phase space where a recoil prescription has been applied.
The technical parameter $\epsilon$ can be pushed to arbitrary low values \mbox{$\epsilon > 0$} since the integral is finite and the cancellation between the two terms is local in $r'$.
Thus, no large numerical cancellations appear after the integration, at variance with Eq.~\eqref{eq:qtsub}.
The origin of the correction in Eq.\,\eqref{eq:mainlpc} can be understood 
as follows: The first term provides an approximation of the $F$+jet cross section 
below the cutoff, including all singular terms in $q_T$ 
and the \lpc, while the second term 
is the usual subtraction term that removes all the singular contributions.
Hence, what remains in their difference are the \lpc below $r_{\rm cut}$, which 
can be directly added to Eq.\,\eqref{eq:qtsub} in order to correct the $q_T$-subtraction
formula for \lpc. Note that Eq.\,\eqref{eq:mainlpc} can also be 
derived directly from expanding the formula for the 
fixed-order matching of $q_T$-resummation with recoil prescription.

The contribution in Eq.~\eqref{eq:mainlpc} can be straightforwardly added to any numerical code that contains an implementation of the $q_T$-subtraction formalism.\footnote{In principle it can also be useful in the context of NNLO-matched predictions that include a $q_T$-slicing cutoff \cite{Hoche:2014uhw,Alioli:2021qbf}.}
We have implemented this contribution in the \Matrix framework~\cite{Grazzini:2017mhc}  by using a boost from the Collins--Soper rest frame of the colour singlet system~\cite{Collins:1977iv} to the laboratory frame where it has transverse momentum equal to $q_T$~\cite{Catani:2015vma,Grazzini:2015wpa}.\footnote{We have also considered other choices of boosts which yield almost undistiguishable results, in agreement with the observations made in Ref.~\cite{Ebert:2020dfc},the effect being $\mathcal O(r_{\rm cut}^2)$.}
We have then studied the effect of adding this contribution for various setups that suffer from \lpc.
In particular, we have focussed on Drell--Yan production with symmetric and asymmetric cuts, which proceeds through $s$-channel diagrams at Born level, and on on-shell $ZZ$ production, where symmetric cuts are applied on the transverse momenta of the two $Z$ bosons.
Although the fiducial region in $ZZ$ production is usually defined through cuts on the decay products of the two $Z$ bosons, in which case no linear behaviour is observed~\cite{Grazzini:2017mhc}, it is interesting to consider a process which at Born level proceeds through $t$-channel diagrams.
For that case a formal proof of the all-order resummation of \lpc along the lines of Ref.~\cite{Ebert:2020dfc} is less straightforward. However, as we will see, resorting to a recoil prescription allows us to include them since the procedure accounts for the phase-space effects responsible for the appearance of the \lpc.

\begin{figure}[t]
  \begin{tabular}{c}
    \includegraphics[width=0.48\textwidth]{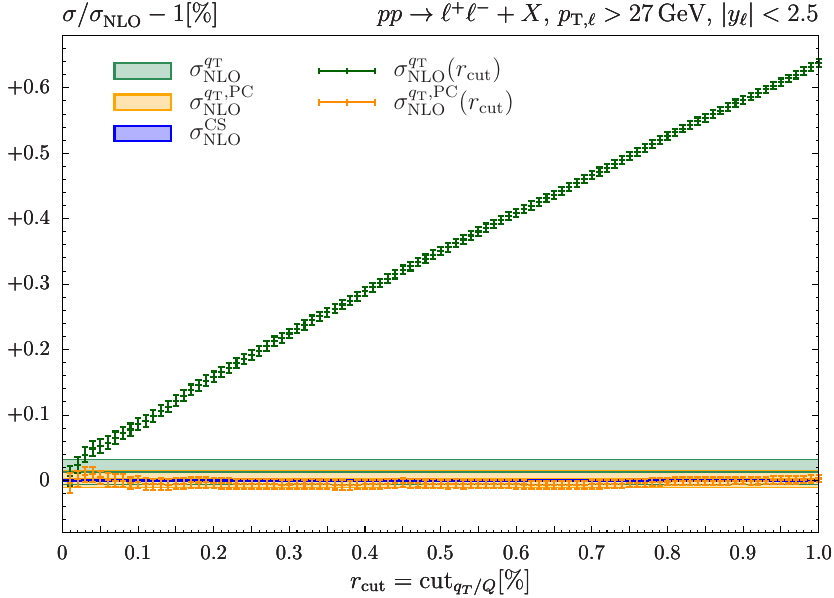}
  \end{tabular}
  \caption{\label{fig:DYsymNLO} Dependence of the NLO QCD Drell--Yan cross section, calculated in the $q_{\mathrm{T}}$-subtraction method with (orange) and without (green) \lpc, on the cutoff $r_{\rm cut}$, normalized to the reference CS result (blue) and with statistical errors. The horizontal lines show the respective \mbox{$r_{\rm cut}\to 0$} extrapolations, with their combined numerical and extrapolation uncertainties depicted as bands.}
\end{figure}

We now turn to discussing the numerical effects of including the \lpc via Eq.\,\eqref{eq:mainlpc} in \Matrix predictions.
Unless stated otherwise, we consider $\sqrt{s}=13$\,TeV proton--proton collisions at the LHC and use the PDF set 
\texttt{NNPDF31\_nnlo\_as\_0118}\,\cite{NNPDF:2017mvq} as well as renormalization and factorization scales $\mu_R=\mu_F = m_Z$.
We start by considering the neutral-current Drell--Yan production process with symmetric cuts on the leptons, requiring a transverse 
momentum of \mbox{$p_{T,\ell}>27$\,GeV} and a rapidity of \mbox{$|y_\ell|<2.5$} for the leptons as well as a \mbox{$66{\rm\,GeV}<m_{\ell\ell}<116$\,GeV}
invariant-mass window for the lepton pair. 
Figure~\ref{fig:DYsymNLO} shows the fiducial cross section at next-to-leading order~(NLO) in QCD as a function of the $q_T$-slicing cutoff $r_{\rm cut}$, normalized 
to the $r_{\rm cut}$-independent reference cross section at NLO QCD that is obtained with Catani--Seymour (CS) subtraction~\cite{Catani:1996vz} in \Matrix, shown in blue. The result without \lpc is 
given in green and that including the \lpc via Eq.\,\eqref{eq:mainlpc} in orange, where the vertical error bars reflect the 
statistical errors for each individual $r_{\rm cut}$ value. The horizontal lines correspond to the \mbox{$r_{\rm cut}\to 0$} extrapolation of the respective cross sections,
using their $r_{\rm cut}$ dependence down to \mbox{$r_{\rm cut}=0.01$\%} by means of the extrapolation procedure 
described in Ref.~\cite{Grazzini:2017mhc}, and the corresponding bands include both numerical and extrapolation uncertainties.

First of all, the extrapolated results in either case are fully consistent with the reference CS prediction. However, it is 
quite remarkable how much the $r_{\rm cut}$ dependence of the cross section reduces once \lpc are included. Indeed, Figure~\ref{fig:DYsymNLO} 
clearly shows that, in case of symmetric cuts on the leptons, the cross section features a linear dependence on the cutoff $r_{\rm cut}$, and that 
this linear dependence is turned into quadratic (at worst) as soon as the \lpc are included.
This observation confirms that \lpc are captured through recoil effects  
as implemented in Eq.\,\eqref{eq:mainlpc}. 
We would like to stress that the remaining effects after including the \lpc are well below
one permille of the NLO QCD cross section. 
While the extrapolated results with and without \lpc are compatible with each other and the reference result, the substantial 
stabilisation of the $r_{\rm cut}$ dependence in the case with \lpc is an important advancement. Within numerical uncertainties essentially any 
fixed $r_{\rm cut}$ value in the plotted range would yield a viable prediction of the cross section such that also comparably high fixed $r_{\rm cut}$ values would provide accurate results.
Using a higher $r_{\rm cut}$ value renders the numerical integration much more efficient since the large cancellations between 
$F$+jet cross section and counterterm in Eq.\,\eqref{eq:qtsub} are significantly reduced. 

Moreover, the \mbox{$r_{\rm cut}\to 0$} extrapolation is fully compatible with the results obtained with a finite value of  $r_{\rm cut}$ in all the range considered in the plot. Whilst the extrapolated result (and its error) provides a more robust prediction than those obtained with finite values of $r_{\rm cut}$, the consistency of the results across $r_{\rm cut}$ when \lpc are included is particularly useful for distributions, for which an automated bin-wise extrapolation is supported only from version 2.1 of the \Matrix code (although already used before \cite{Grazzini:2017ckn,Catani:2019hip,Grazzini:2019jkl,Catani:2020tko, Catani:2020kkl,Kallweit:2020gcp,Buonocore:2021rxx,Bonciani:2021zzf}).

\begin{figure}[t]
  \centering
  \begin{tabular}{c}
    \includegraphics[width=0.48\textwidth]{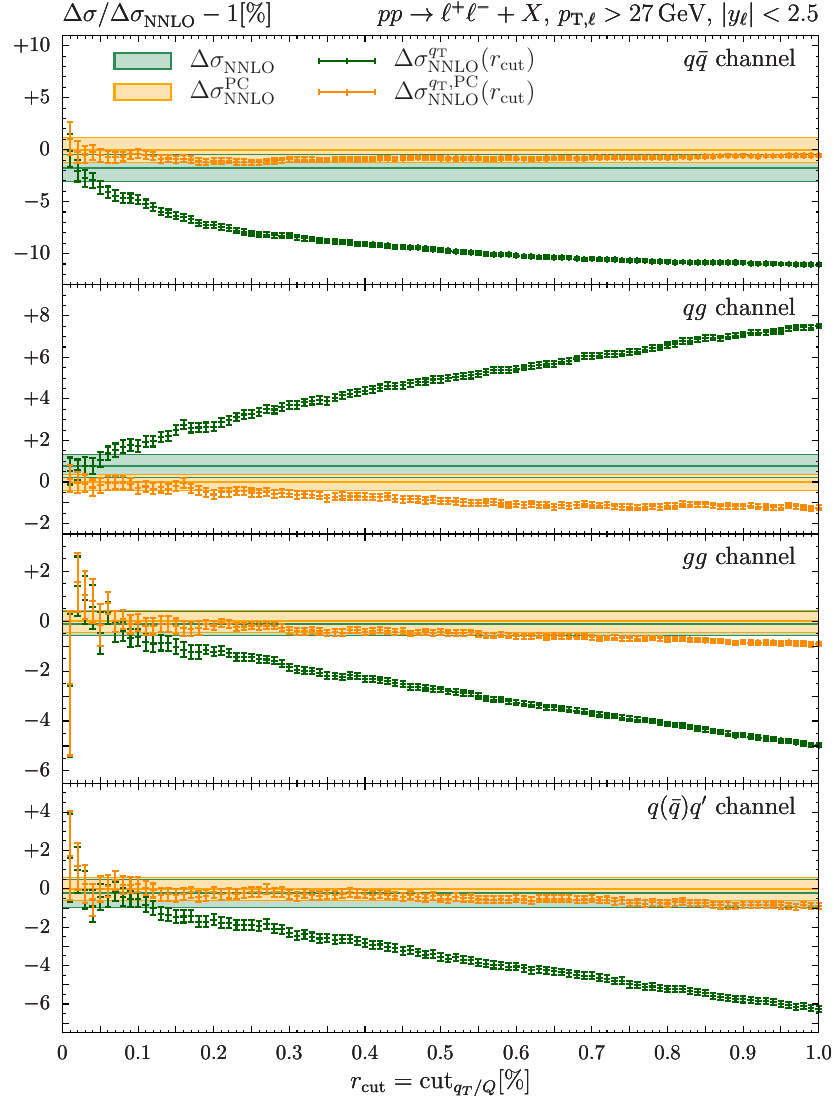} 
  \end{tabular}
  \caption{\label{fig:DYsymNNLO} Dependence of the NNLO QCD Drell--Yan coefficient on $r_{\rm cut}$ for each partonic channel with (orange) and without (green) \lpc, normalized to the \mbox{$r_{\rm cut}\to 0$} result with \lpc. The horizontal lines show the respective \mbox{$r_{\rm cut}\to 0$} extrapolations. Errors indicated as in Figure~\ref{fig:DYsymNLO}. }
\end{figure}

\begin{figure}[t]
  \centering
  \begin{tabular}{c}
    \includegraphics[width=0.48\textwidth]{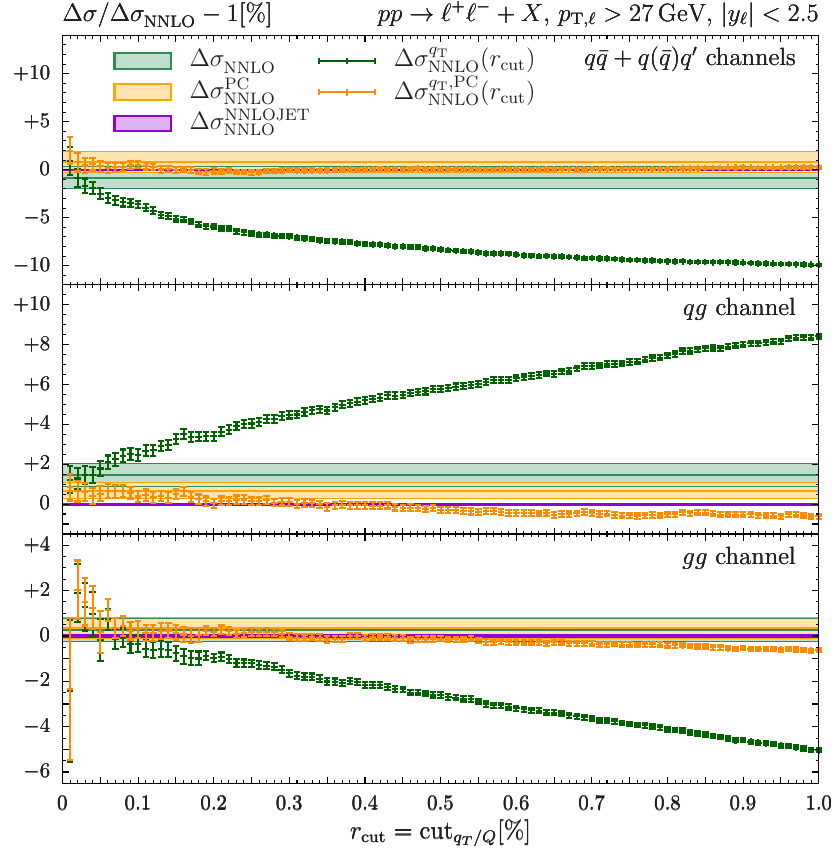} 
  \end{tabular}
  \caption{\label{fig:DYsymNNLO_NNLOjet} Dependence of the NNLO QCD Drell--Yan coefficient on $r_{\rm cut}$ for different partonic channels with (orange) and without (green) \lpc, normalized to the NNLOjet result (purple). The horizontal lines show the respective \mbox{$r_{\rm cut}\to 0$} extrapolations. Errors indicated as in Figure~\ref{fig:DYsymNLO}.}
\end{figure}

While the NLO QCD results presented so far are instructive to study the effects of \lpc in comparison to a reference prediction,
the inclusion of \lpc in the $q_T$-slicing cutoff becomes much more relevant at next-to-NLO (NNLO) in QCD perturbation theory.
The evaluation of the $\mathcal{O}(\as^2)$ coefficient in \Matrix relies entirely on the $q_T$-subtraction method,
 and no $r_{\rm cut}$-independent NNLO QCD cross section can be computed with the code. 
In Figure~\ref{fig:DYsymNNLO} we study the $r_{\rm cut}$ dependence 
of the NNLO QCD coefficient for different partonic channels, normalized to the respective \mbox{$r_{\rm cut}\to 0$} results with \lpc.
The symbols for the partonic channels ($q\bar q$, $qg$, $gg$, $q(\bar q)q^\prime$) are defined as usually, i.e.\ symmetrically with respect to the beam directions: $gg$ for the gluon--gluon channel, $qg$ including all (anti-)quark--gluon channels, $q\bar q$ referring to the diagonal quark--(anti-)quark channels present already at leading order, and $q(\bar q)q^\prime$ collecting all remaining (anti-)quark--(anti-)quark channels such that the four categories sum up to the full result.

In Figure~\ref{fig:DYsymNNLO} we observe that the NNLO QCD coefficient features an analogous reduction in the $r_{\rm cut}$ dependence when accounting for \lpc by including the contribution of Eq.\,\eqref{eq:mainlpc}. We note that starting from NNLO QCD the linear scaling can be 
enhanced by additional logarithms in $r_{\rm cut}$ (i.e.\ terms of order $r_{\rm cut} \ln^k (r_{\rm cut}) $, $k \in \[1,2\]$), as can be seen from the figures.
Like at NLO QCD the extrapolated \mbox{$r_{\rm cut}\to 0$} results are fully compatible, but 
the cross section with \lpc exhibits a considerably reduced $r_{\rm cut}$ dependence with the advantages discussed above.
In Fig.\ref{fig:DYsymNNLO_NNLOjet} we compare the NNLO correction in different partonic channels with the NNLOjet results~\cite{Bizon:2019zgf,Re:2021con}, which are obtained with the $r_{\rm cut}$--independent antenna subtraction method~\cite{Gehrmann-DeRidder:2005btv,Currie:2013vh}. We use the same setup as discussed above, but we now take $\mu_F=\mu_R=\sqrt{m_{\ell \ell}^2 + q_T^2}$. We observe a very good agreement, down to the $\mathcal O (1 \%)$ level of the NNLO coefficient, in all the partonic channels.

\begin{figure*}[t]
  \centering
  \begin{tabular}{cc}
    \includegraphics[width=0.45\textwidth]{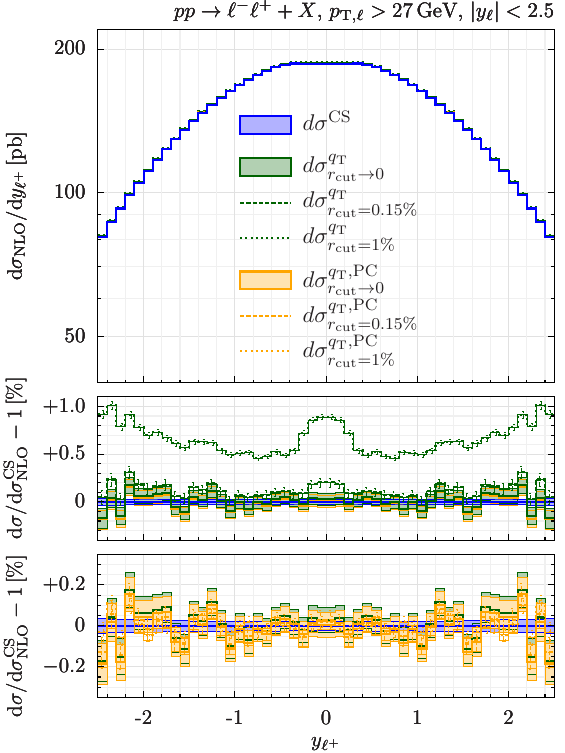} &%
    \hspace{1cm}\includegraphics[width=0.45\textwidth]{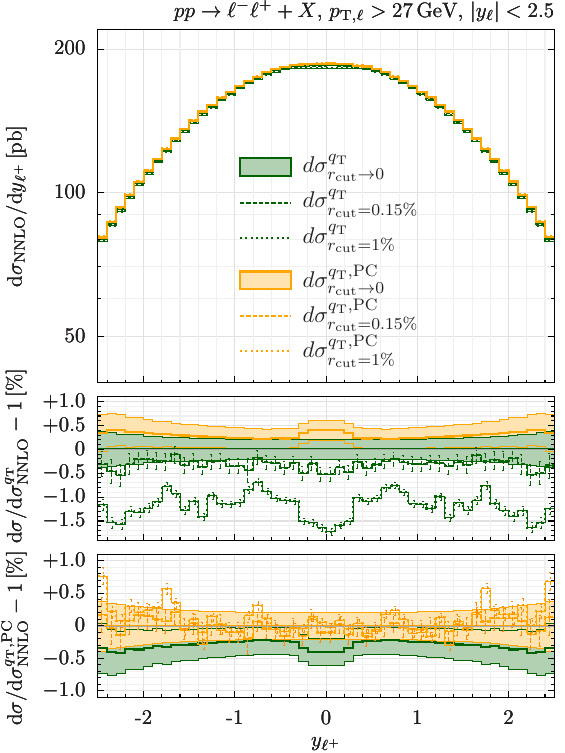} \\
  \end{tabular}
  \caption{ \label{fig:distr} Distribution in the rapidity of the anti-lepton for \mbox{$r_{\rm cut}=1$\%} (dotted), \mbox{$r_{\rm cut}=0.15$\%} (dashed), and \mbox{$r_{\rm cut}\to 0$} (solid with bands), with \lpc (orange) and without (green) at NLO QCD (left) and NNLO QCD (right). For reference, the CS result is shown at NLO QCD (blue) and normalized to in the ratio, while at NNLO QCD the first panel is normalized to the \mbox{$r_{\rm cut}\to 0$} result without \lpc and the second to
    the \mbox{$r_{\rm cut}\to 0$} result with \lpc.}
\end{figure*}
 \begin{figure}[]
     \centering
     \begin{tabular}{c}
    \includegraphics[width=0.45\textwidth]{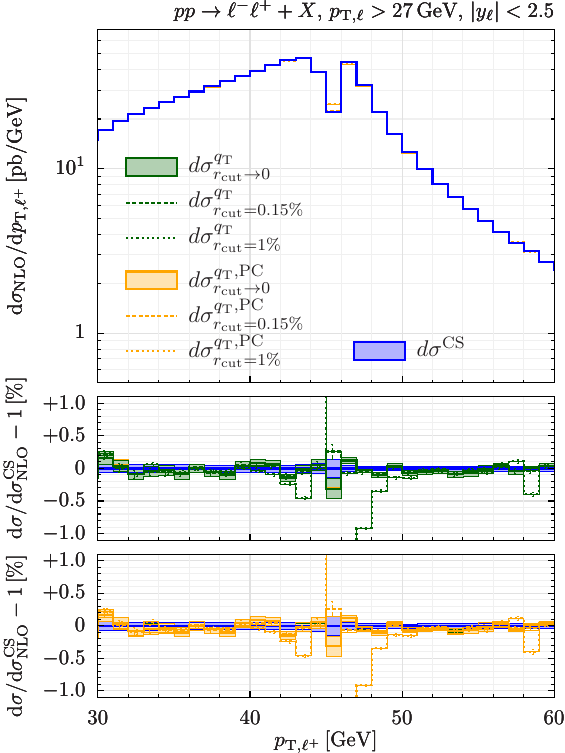} \\
      \end{tabular}
      \caption{ \label{fig:pTl} Same as Figure~\ref{fig:distr}\,(left), but for the transverse momentum of the positively charged lepton.}
 \end{figure}

We continue with the discussion of differential distributions within the fiducial phase-space selection.
Figure~\ref{fig:distr} shows the 
rapidity distribution of the positively charged lepton ($y_{\ell^+}$) at NLO QCD (left) and at NNLO QCD (right) 
in the main panel.
Results for the fixed values \mbox{$r_{\rm cut}=1$\%} (dotted) and \mbox{$r_{\rm cut}=0.15$\%} (dashed) with their statistical uncertainties indicated by error bars are shown with (orange) and without (green) \lpc in the upper and lower ratio panels, respectively.
The extrapolated \mbox{$r_{\rm cut}\to 0$} results with (orange) and without (green) \lpc with their combined numerical and extrapolation uncertainties indicated by bands are depicted in both ratio panels.
At NLO QCD all curves in the two ratio panels are normalized to the reference $r_{\rm cut}$-independent CS result (blue),
 while at NNLO QCD all curves in the upper (lower) ratio panel are normalized to the extrapolated result without (with) \lpc.

The agreement at NLO QCD with the CS result is truly remarkable, especially considering the very fine binning.
As expected, only the curve with a high cutoff (\mbox{$r_{\rm cut}=1$\%}) and without \lpc is off by about $1$\%. 
Notably, this difference at \mbox{$r_{\rm cut}=1$\%} is removed by including the \lpc.
In all cases the extrapolated results are fully compatible with that of the CS calculation at the permille level and within the respective uncertainties.

At NNLO QCD we can appreciate the much better convergence in $r_{\rm cut}$ when \lpc are included. In the first ratio panel, which shows the
curves without \lpc, the \mbox{$r_{\rm cut}=0.15$\%} (\mbox{$r_{\rm cut}=1$\%}) result is about $0.5$\% (more than 1\%) from the extrapolated result.
By contrast, the curves including the \lpc in the second ratio panel all agree within a few permille up to statistical fluctuations. Therefore, the much higher 
$r_{\rm cut}$ value of 1\% would be sufficient to obtain a reliable prediction, which requires substantially less computing time than pushing 
$r_{\rm cut}$ down to very low values to perform a proper extrapolation. We also observe that the extrapolated predictions 
with and without \lpc agree at the level of a few permille, fully covered by the respective uncertainty bands.

\begin{figure*}[t]
  \centering
  \begin{tabular}{cc}
    \includegraphics[width=0.48\textwidth]{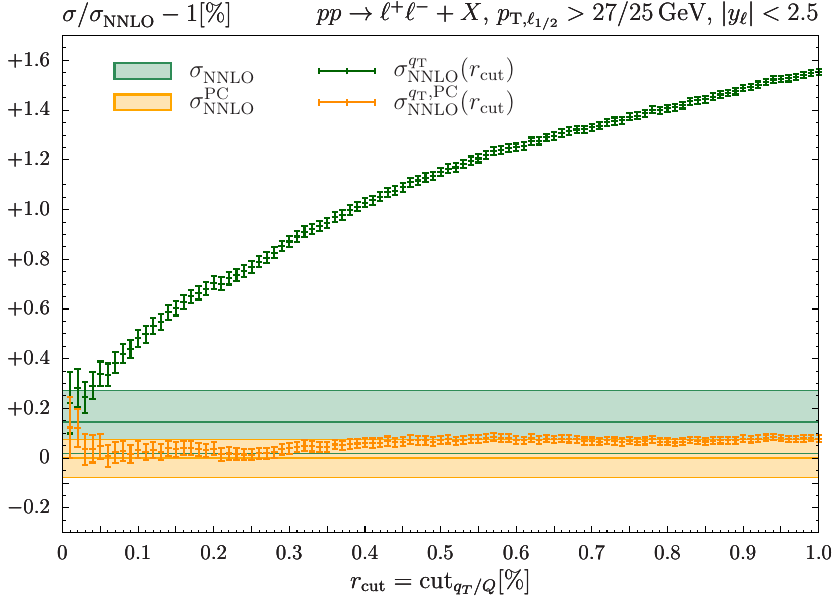} &%
    \includegraphics[width=0.48\textwidth]{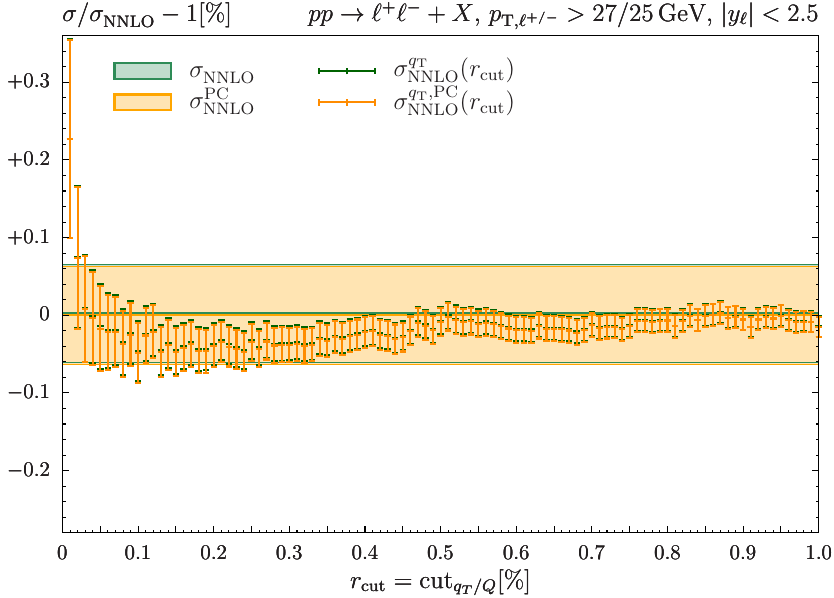} \\
  \end{tabular}
  \caption{ \label{fig:asymstag} Dependence of the NNLO QCD Drell--Yan cross section on $r_{\rm cut}$ with (orange) and without (green) \lpc, normalized to the \mbox{$r_{\rm cut}\to 0$} result with \lpc, for asymmetric cuts (left) and for staggered cuts (right). The horizontal lines show the respective \mbox{$r_{\rm cut}\to 0$} extrapolations. Errors indicated as in Figure~\ref{fig:DYsymNLO}. }
\end{figure*}
 
We have considered various observables of the leptonic final states in Drell--Yan production, and the $y_{\ell^+}$ distribution turned out
to exhibit the largest effects, while similar conclusions can be drawn for all others.
One exception marks, however, the $m_Z/2$ threshold in the transverse-momentum distribution of each lepton ($p_{T,\ell}$), as
shown in Figure~\ref{fig:pTl}, which 
is perturbatively not well-behaved~\cite{Catani:1997xc}. 
Due to the uncancelled large logarithmic contribution in the region \mbox{$p_{T,\ell}\sim m_Z/2$}
the presence of the cutoff $r_{\rm cut}$
causes a discrepancy with a calculation using a local subtraction, which cannot be recovered by applying a recoil prescription,
as observed in Ref.~\cite{Ebert:2020dfc} for charged-current Drell--Yan production.
Since this threshold region can be appropriately described only by a resummed calculation and not at fixed order, 
we do not consider this a drawback
of our approach. We note that the numerical extrapolation \mbox{$r_{\rm cut}\to 0$} exhibits a reasonable convergence 
to a local fixed-order calculation also in that region.

As discussed above, applying asymmetric cuts on the transverse momenta of leading and subleading leptons does not 
cure the issue of \lpc.
We recall that this is a more fundamental problem than just a technical complication for 
slicing approaches, since the linear dependence in $q_T$ ultimately leads to a factorial growth of the coefficients in the perturbative series~\cite{Salam:2021tbm}. 
Only when using staggered cuts, i.e.\ different transverse-momentum thresholds for each individual lepton identified by its charge, 
these problems are avoided entirely.
In Figure~\ref{fig:asymstag} we demonstrate this by showing the NNLO QCD cross sections as functions of $r_{\rm cut}$ for both asymmetric and staggered cuts, normalized to 
the respective \mbox{$r_{\rm cut}\to 0$} results with \lpc. In either case we have kept the same setup as described above, but lowered the transverse-momentum 
threshold for the softer (negatively charged) lepton to $25$\,GeV in the asymmetric-cuts (staggered-cuts) scenario.

 \begin{figure}[t]
     \centering
     \begin{tabular}{cc}
   \includegraphics[width=0.48\textwidth]{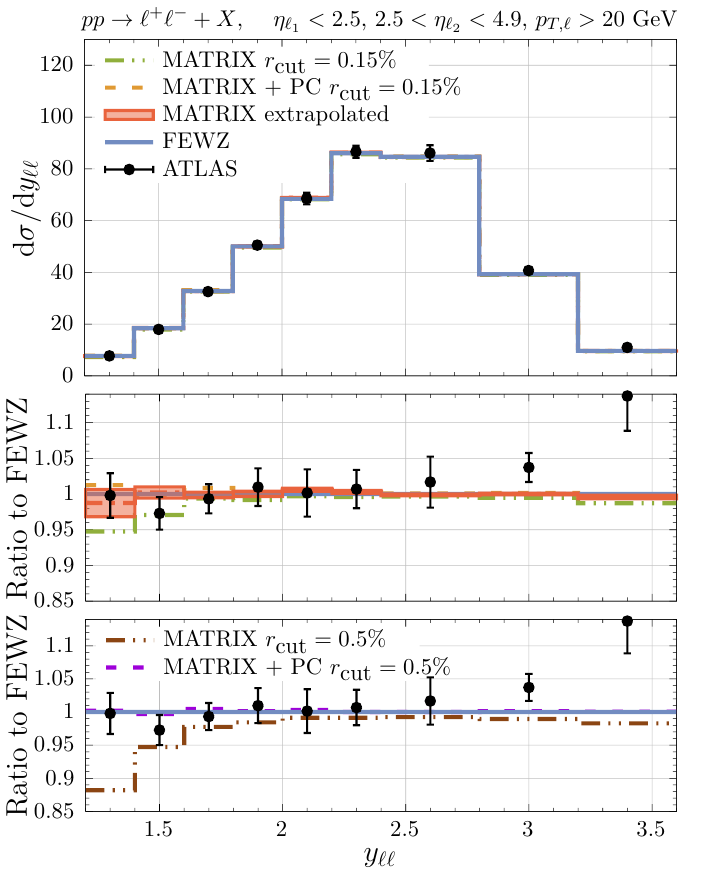}
      \end{tabular}
      \caption{ \label{fig:AKMT} Distribution in the rapidity of the lepton pair for \mbox{$r_{\rm cut}=0.15$\%} without \lpc (green, dash-double-dotted) and its
      extrapolation \mbox{$r_{\rm cut}\to 0$} (red, dash-dotted) as well as with \lpc for \mbox{$r_{\rm cut}=0.15$\%} (orange, dashed) in the setup presented in Ref.~\cite{Alekhin:2021xcu}. For reference, we compare against an $r_{\rm cut}$-independent result by \textsc{FEWZ} (blue, solid) and against ATLAS
      $7$\,TeV data~\cite{ATLAS:2016nqi} (black data points). The first ratio panel shows all results in the main frame normalized to \textsc{FEWZ}, while 
      in the second we show the same ratios, but for results with \mbox{$r_{\rm cut}=0.5$\%} with (purple, dashed) and without (brown, dash-double-dotted) \lpc{}.}
 \end{figure} 
 
We observe the same pattern for asymmetric as for symmetric cuts, with a similarly large and linear $r_{\rm cut}$ dependence (but with opposite sign) without \lpc and a significant reduction
when \lpc are included. On the contrary, the $r_{\rm cut}$ dependence for staggered cuts is completely 
flat, as already pointed out in Ref.~\cite{Grazzini:2017mhc}. In fact, the inclusion of the contribution in Eq.\,\eqref{eq:mainlpc} has practically no 
impact due to the absence of recoil-driven \lpc for staggered cuts.

We stress that the cutoff dependence in $q_T$ subtraction due to missing power corrections is expected to be quadratic in general for QCD corrections to colour singlet production processes~\cite{Ebert:2018gsn,Buonocore:2019puv,Cieri:2019tfv,Oleari:2020wvt}, unless
there are specific fiducial cuts rendering them linear, like for instance symmetric/asymmetric cuts in two-body final states 
or smooth-cone isolation~\cite{Frixione:1998jh} in photon production processes. 
This observation is in line with the findings for single-boson and diboson processes in Ref.~\cite{Grazzini:2017mhc}.
In conclusion, we observe that a difference $\delta p_{T} $ of 2\,GeV between electron and positron transverse-momentum thresholds is sufficient to 
eliminate the linear dependence. This is in line with an explicit calculation of power corrections in the fiducial acceptance, which shows that, in most of the phase space, linear power corrections are absent as long as $q_T < \delta p_{T} $~\cite{Alekhin:2021xcu,Salam:2021tbm}.
Due to the aforementioned instabilities related to the presence of a linear dependence in $q_T$, staggered cuts constitute a feasible option for future analyses, alongside alternative cuts~\cite{Salam:2021tbm} that we did not consider here.

Next, we would like to add few comments on the results shown in Ref.~\cite{Alekhin:2021xcu} about the intrinsic 
uncertainties of non-local subtraction methods for the computation of higher-order corrections in Drell--Yan production.
Given the particularly high precision of Drell--Yan measurements and the resulting 
demand for very accurate theory predictions, full control
on the systematic uncertainties associated to $q_T$ subtraction is highly desirable.
This is crucial not only in the 
context of NNLO QCD corrections, but also for recent developments of computing next-to-NNLO (N$^3$LO) cross sections
using $q_T$ subtraction~\cite{Cieri:2018oms,Chen:2021vtu,Billis:2021ecs,Camarda:2021ict}.

While \Matrix results (at fixed $r_{\rm cut}$) are within about $1\%$ of those obtained with~\textsc{FEWZ}~\cite{Gavin:2010az}, for most neutral-current Drell--Yan production distributions shown in Ref.~\cite{Alekhin:2021xcu}
they deviate from \textsc{FEWZ} by  a few-percent in the first two bins
of the dilepton rapidity ($y_{\ell\ell}$) distribution in a very specific setup, shown in the rightmost plot of Figure~6 of that paper.
In this setup, the harder lepton is in the central rapidity region (\mbox{$|y_{\ell_1}|<2.5$}), while the softer 
is forward in rapidity (\mbox{$2.5<|y_{\ell_2}|<4.9$}), in addition to the standard requirements \mbox{$p_{T,\ell}>20$\,GeV} and \mbox{$66{\rm\,GeV}<m_{\ell\ell}<116$\,GeV}.
In Figure~\ref{fig:AKMT} we repeat the comparison done in Ref.~\cite{Alekhin:2021xcu} using the same setup, namely
$\sqrt{s}=7$\,TeV and \texttt{ABMP16\_5\_nnlo}\,\cite{Alekhin:2017kpj} PDFs with $\as(m_Z)=0.1147$. We include the following predictions: \Matrix at fixed \mbox{$r_{\rm cut}=0.15$\%} (green, dash-double-dotted), 
the corresponding \mbox{$r_{\rm cut}\to 0$} extrapolation (red, dash-dotted), our novel \Matrix predictions 
with \mbox{$r_{\rm cut}=0.15$\%} including \lpc (orange, dashed), and, as a reference, the prediction obtained with \mbox{\textsc{FEWZ}} (blue. solid) as well as  
$7$\,TeV ATLAS data (black, with error bars).\footnote{We would like to thank the authors of Ref.~\cite{Alekhin:2021xcu} for providing us with the \textsc{FEWZ} results of Figure~6 in Ref.~\cite{Alekhin:2021xcu}.}
In the first ratio panel
all results of the main frame are shown normalized to \mbox{\textsc{FEWZ}}. In the lower panel corresponding ratios
for \mbox{$r_{\rm cut}=0.5$\%} with (purple, dashed) and without (brown, dash-double-dotted) \lpc{} can be appreciated.

Using a fixed value of \mbox{$r_{\rm cut}=0.15$\%} without \lpc results in differences up to $\sim 5$\% with respect to 
the \mbox{\textsc{FEWZ}} prediction in the first two bins, as already shown in Ref.~\cite{Alekhin:2021xcu}.
Indeed, those may be considered too large
for current precision studies of the Drell--Yan process, 
although the $7$\,TeV ATLAS errors cannot resolve these differences.
The inclusion of the \lpc is sufficient to 
obtain agreement with \mbox{\textsc{FEWZ}} within 1\% at an \mbox{$r_{\rm cut}=0.15$\%}.
Increasing to a fixed $r_{\rm cut}$ value of 0.5\%
makes the comparison even more striking, as shown in the lower ratio panel:
The discrepancy to the \mbox{\textsc{FEWZ}} results in the first bins is further increased without \lpc, whereas the agreement is excellent throughout
as soon as they are included.

From the first ratio panel in Figure~\ref{fig:AKMT} we observe that 
the \mbox{$r_{\rm cut}\to 0$} extrapolation is sufficient for 
the \Matrix prediction to become compatible with that of \mbox{\textsc{FEWZ}} within 1\%, which is covered by the quoted error band that includes
both statistical and extrapolation uncertainties. One has to bear in mind, however, 
that the \mbox{$r_{\rm cut}\to 0$} extrapolation before version 2.1 of \Matrix could 
be obtained only by performing separate runs for each 
bin in a distribution. The support for a bin-wise extrapolation
is available from version 2.1 of \Matrix.
 The previous observations manifest the clear advantage of the approach presented in this letter 
for configurations dominated by a recoil-driven linear cutoff dependence:
The inclusion of \lpc allows one to perform the extrapolation procedure at higher values of $r_{\rm cut}$, without spoiling the accuracy of the calculation. 
This avoids evaluating and storing results down to very small $r_{\rm cut}$ 
values in all bins of differential distributions in order to perform meaningful \mbox{$r_{\rm cut}\to 0$} extrapolations.
Therefore, the numerical computation becomes substantially less demanding, reducing considerably the computing time.
We note that the very good agreement between the results obtained with the recoil
prescription and those obtained using a \mbox{$r_{\rm cut}\to 0$} extrapolation constitute 
a consistency check that the extrapolation is robust in this case. This is an indication of the reliability of the extrapolation procedure, which is 
the only viable strategy for cases in which the linear power corrections have a different origin.
 
 \begin{figure*}[t]
     \centering
     \begin{tabular}{cc}
   \includegraphics[width=0.48\textwidth]{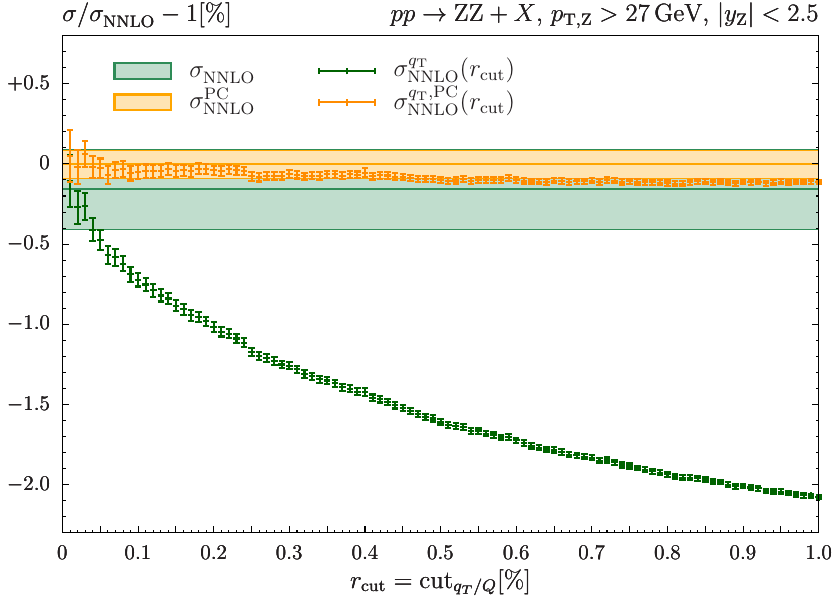} &%
   \includegraphics[width=0.48\textwidth]{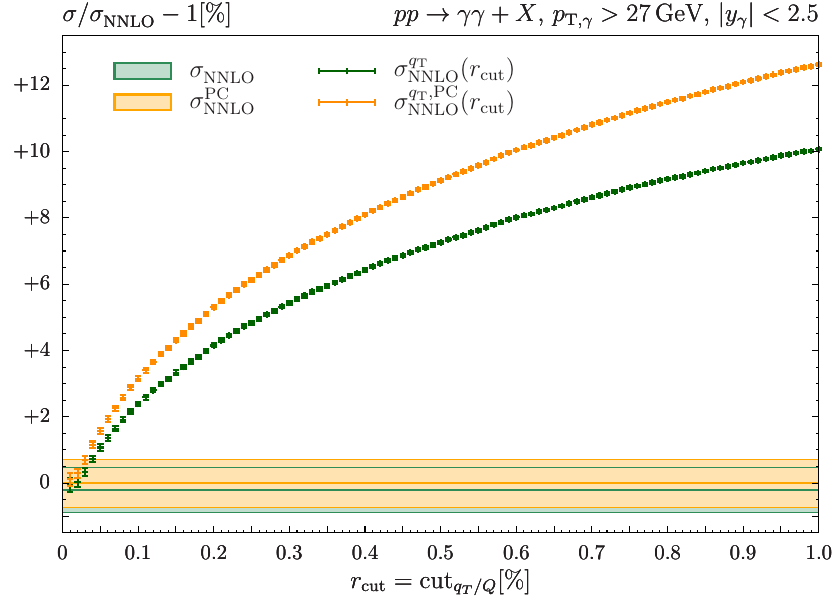} \\
      \end{tabular}
      \caption{ \label{fig:ZZgg} Dependence of the NNLO QCD cross section on $r_{\rm cut}$ with (orange) and without (green) \lpc, normalized to the \mbox{$r_{\rm cut}\to 0$} result with \lpc, for $ZZ$ production (left) and for $\gamma\gamma$ production (right).  The horizontal lines show the respective \mbox{$r_{\rm cut}\to 0$} extrapolations. Errors indicated as in Figure~\ref{fig:DYsymNLO}. For $\gamma\gamma$ production $\mu_R=\mu_F = \sqrt{m_{\gamma\gamma}^2+p_{T,\gamma\gamma}^2}$ is used.}
 \end{figure*}

 \begin{figure}[t]
     \centering
     \begin{tabular}{c}
   \includegraphics[width=0.48\textwidth]{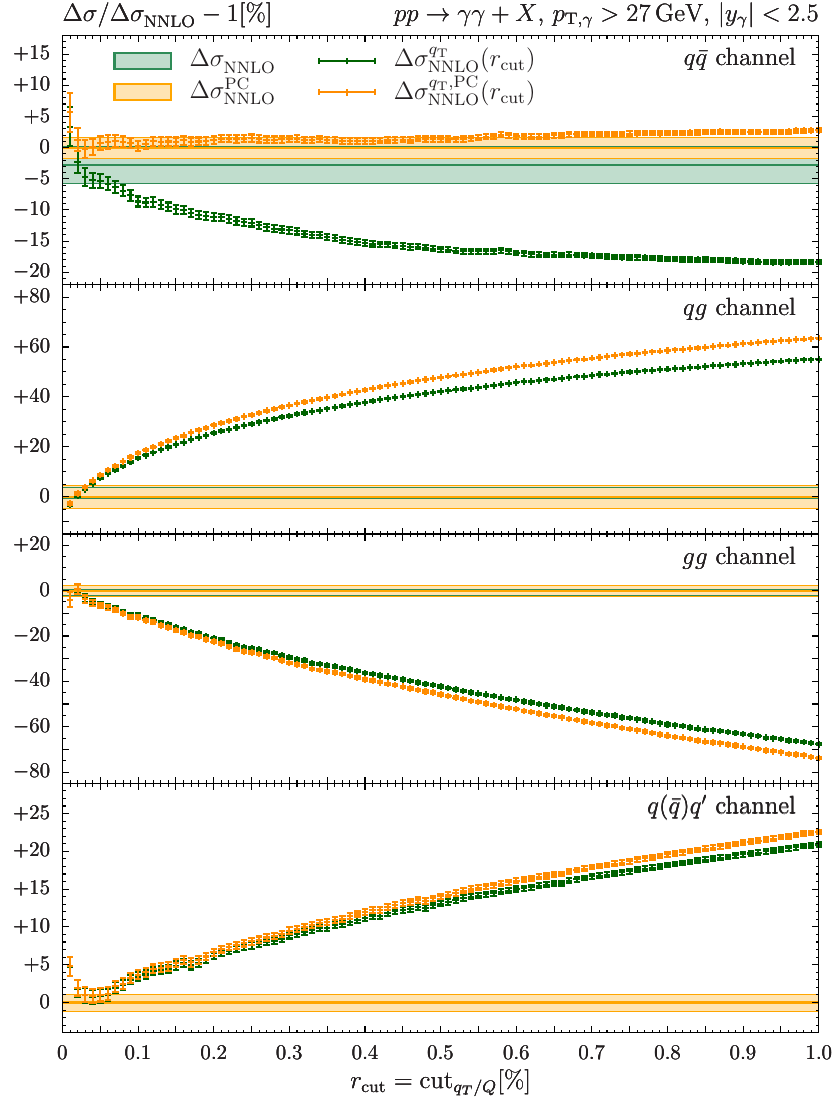} \\
      \end{tabular}
      \caption{ \label{fig:qqdiphoton} Dependence of the NNLO QCD coefficient for $\gamma\gamma$ production on $r_{\rm cut}$ for each partonic channel with (orange) and without (green) \lpc, normalized to the \mbox{$r_{\rm cut}\to 0$} result with \lpc.  The horizontal lines show the respective \mbox{$r_{\rm cut}\to 0$} extrapolations. Errors indicated as in Figure~\ref{fig:DYsymNLO}.}
 \end{figure}
 
Finally, we have also considered other processes with two-particle final states, in particular on-shell $ZZ$ and $\gamma\gamma$ production. 
For $ZZ$ production it has already been shown that power corrections in 
the inclusive case or in a usual fiducial setup with cuts on the four-lepton final state
in off-shell $ZZ$ production
are relatively flat and have the expected quadratic dependence on $r_{\rm cut}$~\cite{Grazzini:2017mhc}.
Therefore, we have chosen 
a non-standard set of fiducial cuts that impose symmetric cuts on the two on-shell $Z$ bosons.
This provides an interesting sample case
since on-shell $ZZ$ production proceeds through $t$-channel diagrams at Born level and
the formal proof~\cite{Ebert:2020dfc} for the resummation of \lpc in Drell--Yan production does not directly generalise to the $ZZ$ process.
Thus, we study here whether \lpc for the $ZZ$ process with symmetric $Z$-boson cuts exist and can be described by suitably 
accounting for the recoil in $q_T$ subtraction through Eq.\,\eqref{eq:mainlpc}. By contrast, for $\gamma\gamma$ production
it is well known~\cite{Grazzini:2017mhc,Ebert:2019zkb,Becher:2020ugp} that, as for any process with 
identified photons in the final state, power corrections are linear due to the requirement of 
consistently defining isolated photons through smooth-cone isolation.  The possibility to single out the linear power corrections due to the presence of symmetric cuts and those induced by the isolation requirements allows us to investigate whether there is any hierarchy between the size of the linear power corrections of different origin in a realistic setup. In particular, here we shall study whether including 
recoil effects through Eq.\,\eqref{eq:mainlpc} yields any improvements in the diphoton case.

Figure~\ref{fig:ZZgg} shows the NNLO QCD cross section as a function of $r_{\rm cut}$ normalized to 
the \mbox{$r_{\rm cut}\to 0$} result with \lpc for both $ZZ$ and $\gamma\gamma$ production. 
The symmetric cuts are inspired by the lepton cuts we applied in the case of Drell--Yan production, i.e.\ we have imposed 
a transverse-momentum cut of \mbox{$p_{T,V}>27$\,GeV} and a rapidity requirement of \mbox{$|y_V|<2.5$} on each vector boson \mbox{$V\in\{Z,\gamma\}$}.
Indeed, we observe a linear dependence on $r_{\rm cut}$ also for $ZZ$ production with symmetric cuts, and the \lpc are completely 
included by the contribution of Eq.\,\eqref{eq:mainlpc}, which properly accounts for recoil effects, also in this case.

For diphoton production, on the other hand, the situation is very different. The observed power corrections 
are extremely large for the given setup, even larger than for the setup considered in Ref.~\cite{Grazzini:2017mhc}. 
It is worth noting that the recoil-driven \lpc are independent of those due to photon isolation. In fact, with symmetric cuts the inclusion of recoil-driven \lpc does actually even slightly increase the $r_{\rm cut}$ dependence, whereas the opposite 
behaviour is found when considering asymmetric cuts on leading and subleading photon (not shown here).
This shows that a recoil prescription is not suitable to account for the dominant $r_{\rm cut}$ dependence in
processes with isolated photons, which was also observed in Ref.~\cite{Becher:2020ugp} in the context of
transverse-momentum resummation of the diphoton pair in a different fiducial region.

Nevertheless, it is interesting to notice that the observed behaviour for $\gamma\gamma$ production 
depends on the partonic channel under consideration. In the $q\bar q$ channel, including recoil effects through Eq.\,\eqref{eq:mainlpc} 
is sufficient to account for the \lpc, as shown in Figure~\ref{fig:qqdiphoton}, which is true both at NLO QCD and NNLO QCD.
For all other partonic channels this is not the case and 
the qualitative behaviour is similar to  that observed for their sum in Figure~\ref{fig:ZZgg}\,(right).
The fact that the recoil prescription is sufficient to include the \lpc for the $q\bar{q}$ channel at NLO can be understood as follows:
Problematic configurations in the photon smooth-cone isolation are those where a light quark is close to a photon, as  
collinear photon emissions from quarks lead to QED singularities. Such effects do not appear in the $q\bar{q}$ channel
up to NLO QCD. On the other hand, at NNLO QCD the only configurations that lead to QED singularities and contribute at small $q_T$
are double-real corrections in which both extra emissions become collinear to the emitted photons balancing each other. A possible explanation for 
the absence of linear power corrections at NNLO when including the recoil can be related to the fact that these configurations are however
particularly symmetric. 
 The interplay between the recoil procedure and the isolation requirements is therefore intrinsically different 
in this channel with respect to the others. 
Moreover, those configurations could simply be sufficiently suppressed by phase space, and, in fact, 
such configurations are removed below $r_{\rm cut}$ in a $q_T$-subtraction computation for any process.
A rigorous explanation of this interesting feature characterising the $q \bar q$ channel requires further studies, which we leave to future work.

In this letter, we have presented a relatively simple approach to include linear power corrections 
in fixed-order calculations obtained with slicing methods. This is the first time such corrections are 
included in $q_T$ subtraction for general colour-singlet processes.
Our approach is applicable whenever the 
linear power corrections are of kinematical origin and can thus be captured through an appropriate recoil
prescription. This is the case  if a common transverse-momentum requirement is applied  
on each particle of a process with (effective) two-body kinematics, or  if different transverse-momentum requirements
are applied, but on the undistinguished particles ordered in transverse momentum. We have shown for the case of neutral-current Drell--Yan 
production that such symmetric or asymmetric cuts applied on the leptons lead to a linear dependence 
on the $q_T$-slicing cutoff, and that by following the approach suggested in this letter those linear power 
corrections are accounted for, both at the level of fiducial cross sections and differential distributions.

We have also addressed the concerns raised in Ref.~\cite{Alekhin:2021xcu} about the intrinsic 
uncertainties of differential Drell--Yan predictions in $q_T$ subtraction. Given the enormous precision of Drell--Yan 
studies at the LHC, these concerns are justified when predictions with only a fixed $q_T$-slicing cut are used.
Our suggested approach to include the linear power corrections alleviates these issues even when 
a fixed value of the cutoff is used.
We also observed that it is sufficient to perform a suitable extrapolation of the 
$q_T$-slicing cutoff to zero with \Matrix. 
The latter, however, requires
considerably more computing resources to reach an analogous numerical precision.

Finally, we have considered both $ZZ$ and $\gamma\gamma$ production with 
symmetric transverse-momentum thresholds on the vector bosons and showed that for $ZZ$ production 
the resulting linear power corrections are fully captured by our approach. On the contrary, for $\gamma\gamma$
production such procedure is insufficient, since the need for isolating the photons 
yields an additional source of linear power corrections, which can not be captured through recoil effects.

We have implemented the approach presented here within the \Matrix framework. The additional 
contribution that includes the linear power corrections induced by recoil effects can be turned on 
separately in the input files of all \Matrix processes. This feature is included 
 in the public \Matrix framework from version 2.1. We consider it a useful feature especially for experimentalists that are
interested in obtaining predictions for Drell--Yan production with \Matrix, which  provides both
NNLO QCD and NLO EW corrections, as well as mixed QCD--EW corrections to be included in a future release.
However, while in particular for legacy Drell--Yan analyses the inclusion of the relevant power corrections
is crucial, we recommend to avoid the issues related to the enhanced sensitivity to low momentum scales 
by imposing different sets of cuts in future analyses. As we have 
shown, for staggered cuts a difference of $\mathcal O(\rm GeV)$  between the transverse momentum thresholds
of the individual leptons identified by their charges
is already sufficient to avoid a linear dependence in $q_T$in the relevant $r_{\rm cut}$ range for the computation of higher-order corrections.
 \\[0.5cm]
 {\bf Note added.} An equivalent method to include linear fiducial power corrections in the $q_T$-subtraction formalism has been contemporarily presented in Ref.~\cite{Camarda:2021jsw}.
 \\[0.5cm]
 {\bf Acknowledgements.}  We are indebted to  
 Markus Ebert and Massimiliano Grazzini for fruitful discussions and
 comments on the manuscript. We are grateful to the authors of Ref.~\cite{Alekhin:2021xcu} for helpful 
 correspondence and for sending us some of the numbers used in their study.
 We thank A.~Huss for providing the NNLOJET results. 
 We would like to thank Pier Monni, Emanuele Re, and Paolo Torrielli for discussion on this topic. 
 We also acknowledge discussions about the feasibility of using staggered cuts in experimental
 analyses with Josh Bendavid, Lorenzo Bianchini, and Luigi Rolandi.
 LB and LR are supported by the Swiss National Science Foundation (SNF) under contract 200020\_188464,
 while LB is also supported  in part by the UZH Postdoc Grant Forschungskredit K-72324-03.
 The work of SK is supported by the ERC Starting Grant 714788 REINVENT.

\bibliographystyle{apsrev4-1}
\bibliography{pw_qt_recoil.bib} 

\clearpage

\onecolumngrid
\newpage
\appendix

\end{document}